\documentclass[twocolumn,aps,prc,showpacs,preprintnumbers,amsmath,amssymb]{revtex4}
\usepackage{epsfig} 
\usepackage{dcolumn}

\newcommand{\fourplusfour}{$^4$He + $^4$He}
\newcommand{\fourplusthree}{$^4$He + $^3$He}
\newcommand{\threeplusfour}{$^3$He + $^4$He}

\begin{document}
\title{ The \boldmath{$^3$He + $^4$He $\rightarrow$ $^7$Be} Astrophysical \boldmath{$S$}-factor} 

\author{T. A. D. Brown}
\author{C. Bordeanu}
\altaffiliation{present address: HH-NIPNE, Bucharest, Romania.}
\author{K. A. Snover}
\email{snover@u.washington.edu}
\author{D. W. Storm}
\author{D. Melconian}
\author{A. L. Sallaska}
\author{S. K. L. Sjue}
\author{S. Triambak}
\affiliation{Center for Experimental Nuclear Physics and Astrophysics, University of  Washington,
Seattle,~Washington~98195}

\date{\today}

\begin{abstract}

We present precision measurements of the  $^3$He + $^4$He $\rightarrow$ $^7$Be reaction in the range $E_{\rm c.m.} = 0.33$ to 1.23~MeV using a small gas cell and detection of both prompt $\gamma$ rays and $^7$Be activity.  Our prompt and activity measurements are in good agreement within the experimental uncertainty of several percent.  We find ${S}(0) = 0.595\pm0.018$~keV\,b from fits of the Kajino theory to our data.   
We compare our results with published measurements, and we discuss the consequences for Big Bang Nucleosynthesis and for solar neutrino flux calculations. 

\end{abstract}

\maketitle

\section{Introduction}

The $^3$He + $^4$He $\rightarrow$ $^7$Be fusion reaction is an important step in the solar p-p chain responsible for producing solar neutrinos from  $^7$Be and $^8$B decay.  Bahcall and Pinsonneault~\cite{bahcall} identified it as the solar fusion reaction rate most in need of further study, due to its relatively large uncertainty.  It is also reponsible for essentially all of the $^7$Li produced in the Big Bang, a nuclide whose apparent primordial abundance remains a puzzle~\cite{bbn}.

\begin{figure}
\includegraphics[width=0.45\textwidth]{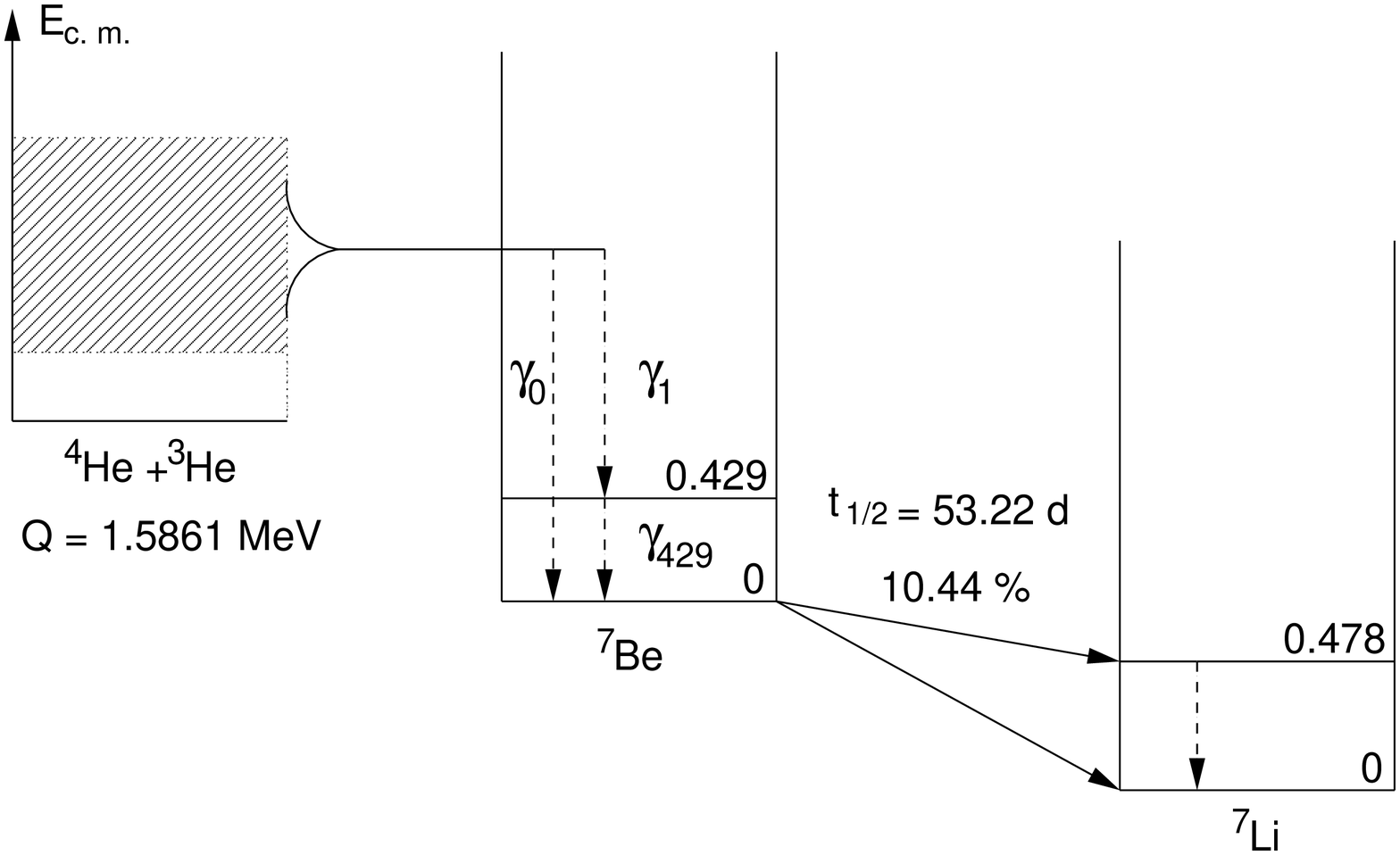}
\caption{Energy level diagram.}
\label{energylevel}
\end{figure}

The $^3$He + $^4$He $\rightarrow$ $^7$Be reaction rate has been determined previously by detecting the prompt $\gamma$ rays~\cite{holm,park,naga,kraw,osbo,alex,hilg} and by counting the $^7$Be activity~\cite{osbo,robe,volk} (see Fig.~\ref{energylevel}).  The accepted value for the zero-energy $S$-factor for this reaction is based on a 1998 recommendation of $S(0) = 0.53 \pm 0.05$~keV\,b~\cite{rmp}, where the relatively large uncertainty stems from an apparent disagreement between the results from the two methods.  Since then, new activity meaurements have been published by a Weizmann Institute group~\cite{nara}, and the LUNA collaboration has presented new activity and prompt results~\cite{gyur,conf}.

Our experiment was designed to measure both the prompt $\gamma$ rays and the $^7$Be activity produced in the same irradiation.  The prompt $\gamma$ yield was measured with a Ge detector at 90$^{\circ}$ with respect to the beam axis.  In order to contain the $^7$Be activity with a high efficiency and accurately define the active volume, we used a $^3$He gas cell target 29.7-mm long with a thin, 1-$\mu$m nickel entrance foil.   Gas pressures of 100 and 200~torr were used.  The copper beam stopper at the end of the gas cell was removed after $^4$He irradiation and counted together with the thin Ta gas cell liner to determine the $^7$Be activity.  We measured the cross section in the range $E_{\rm c.m.} = 0.33$ to 1.23~MeV  with statistical uncertainties of 3\% or less for each data point, and an overall systematic uncertainty of  3\% for the activity data and 3.5\% for the prompt data.

\section{Experimental procedure}
\label{experimentalprocedure}

\subsection{Experimental apparatus and procedure}
\begin{figure}
\includegraphics[width=0.45\textwidth]{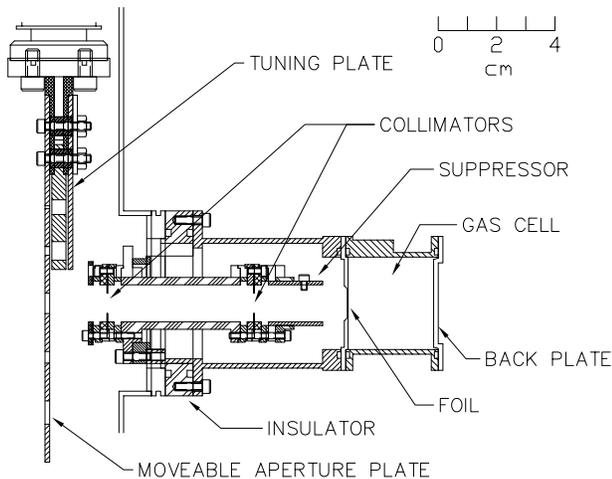}
\caption{Gas cell and beam collimation geometry.}
\label{cell}
\end{figure}

We measured the \threeplusfour\ fusion cross sections using the Model FN Tandem Van de Graaff accelerator at CENPA, the Center for Experimental Nuclear Physics and Astrophysics at the University of Washington.  $^4$He$^+$ beams of $1.5 - 3.5$ MeV were produced in an rf-discharge ion source mounted in the terminal of the accelerator.  The beam current on the target cell was typically  $450 - 500$~nA, which, together with beam rastering (see below), was designed to achieve good entrance foil lifetime.  Between the switching magnet and the $^3$He gas target, the beam passed through two magnetic quadrupole doublets, a set of X--Y magnetic rastering coils and two LN$_2$ traps.  A 7-mm water-cooled aperture was located just downstream of the second  LN$_2$ trap.  A second water-cooled 7-mm aperture was located 17~cm further downstream on a moveable aperture plate, shown on the left side of Fig.~\ref{cell}.    The beam then passed through two 8-mm diameter cleanup collimators, a cylindrical electron suppressor, the 1-$\mu$m nickel entrance foil mounted on a nickel foil holder (with a 10-mm diameter aperture) and into the cell as shown in Fig.~\ref{cell}.  The apertures and collimators were made from Oxygen-Free High Conductivity (OFHC) copper and the beamline and the region immediately upstream of the gas cell were pumped by cryopumps.

The beam was tuned by focussing through a 1-mm diameter aperture located on the moveable aperture plate and backed by a tuning beam stop (see Fig.~\ref{cell}).  Then the beam was rastered to a square distribution somewhat larger than $7 \times7$~mm using 19-Hz and 43-Hz X and Y raster frequencies, and sent through the 7-mm diameter moveable aperture into the gas cell.    

The cell consisted of an aluminum cylinder with a 32-mm inside diameter, a 29.7-mm active length, and a 30-mm diameter, 0.25-mm thick OFHC copper beam stop soldered onto a copper back plate.  The choice of OFHC copper for the beam stop was motivated by its low prompt background, low Li and B contamination, and small  backscattering probability for low energy  $^7$Be ions.  The back plate was air-cooled, and the cell wall was lined with 0.025-mm thick Ta foil in order to catch $^7$Be recoils that did not end up in the stopper.  
 
Prompt $\gamma$ rays were detected in an N-type Ge detector with 100\% relative efficiency  located at 90$^{\circ}$ with its cryostat front face $\approx5$~cm from the center of the cell.  A 3.2-mm thick Plexiglass absorber was placed between the cell and the detector to ensure that $\beta$ particles in coincidence with $\gamma$ rays from calibration sources did not penetrate into the Ge.  The Ge detector was uncollimated, and shielded on the sides by 20~cm of Pb and on the rear by a partial 10-cm Pb shield.  We determined the total prompt $^3$He($^4$He,$\gamma$)$^7$Be  cross section from the $\gamma_0$ yield  and the relatively sharp, isotropic 429-keV secondary following $\gamma_1$ emission.  We used the $\gamma_0$ photopeak centroid to determine the mean reaction energy.  \fourplusfour\ background measurements were made at each bombarding energy in order to determine the prompt $\gamma$-ray background and provide a check on possible contaminant $^7$Be production.   

After irradiation, the stopper was removed from the back plate using a knife, and the delayed 478-keV $\gamma$ rays from $^7$Be decay in the stopper and the Ta liner were counted together in a close geometry using a second (offline) 100\% Ge detector in an enclosed 20-cm thick Pb shield.

\subsection{Ge calibration}
The Ge detectors were calibrated using  $^{54}$Mn, $^{60}$Co, $^{133}$Ba, $^{137}$Cs,  $^{88}$Y, $^{113}$Sn and $^{203}$Hg sources with activities in the range $1 - 7$~kBq specified to an accuracy of $\pm(0.7 - 1.2)\%$ (1$\sigma$)~\cite{IP}.   We also used $^7$Be and $^{24}$Na sources made by the $^{10}$B(\emph{p},$\alpha$)$^7$Be  and $^{23}$Na(\emph{d,p})$^{24}$Na reactions.  {\sc Penelope}~\cite{penelope} Monte Carlo simulations were used to calculate finite source size correction factors for the online detector.  The source line intensities were determined from asymmetric Gaussian function fits to the photopeaks.   The energy calibration of the online detector was determined using \textit{in situ} naturally occuring $\gamma$-ray lines, particularly the 1460.85- and 2614.55-keV lines from  $^{40}$K and $^{208}$Tl decay.

\subsubsection{Offline (activity) Ge detector}
\label{offline-ge}

The offline Ge detector efficiency was measured in two ways.  First, the $^7$Be source strength was determined using the online Ge detector together with the calibrated sources listed above, at a source-to-detector distance of 25~cm, large enough that $\gamma - \gamma$ summing was unimportant.  Then the $^7$Be source was used to calibrate the offline Ge in the close geometry, approximately 2.7~mm from the front face of the carbon fiber cryostat window.  Second, the $^{54}$Mn, $^{113}$Sn and $^{203}$Hg calibration sources were used to calibrate the close geometry.  The results of these two methods agreed well, with an absolute efficiency determination at 478~keV of typically $0.1246 \pm 0.0015$.  A measured $\pm0.25$-mm longitudinal variation in source and stopper placement resulted in an additional $\pm1\%$ systematic error for the different measurements based on a measured efficiency sensitivity of 4\% per mm displacement.  A {\sc trim}~\cite{trim} calculation indicated that the spatial profile of the $^7$Be implanted in the stopper was approximately a  9-mm diameter disk.  We included a 0.997 correction factor for this effect based on off-axis $^7$Be source efficiency measurements.

{\sc trim} calculations for our beam energies, foil and gas thicknesses indicated that $>99\%$ of the $^7$Be atoms produced by $^3$He + $^4$He $\rightarrow$ $^7$Be should be implanted in the beam stop.   Since Be adheres readily to surfaces, any remaining $^7$Be should have deposited on the interior surfaces of the cell, of which 80\% was represented by the beam stop plus the Ta liner.

We counted the $^7$Be activity with the stopper mounted as described above, and the Ta liner cut up and mounted directly behind the stopper. In this geometry the $\gamma$ rays from the liner were counted with an efficiency reduced by  5\% due to the greater source-to-detector distance as well as absorption in the stopper and Ta.  Separate measurements for several bombarding energies in which only the Ta liner was counted indicated $<2\%$ for the fraction of $^7$Be activity on the liner;  thus the reduced efficiency for counting the liner resulted in a negligible additional uncertainty.  Our overall systematic uncertainty in the efficiency for counting the $^7$Be activity was $\pm1.6\%$.   We assumed a $^7$Be half-life of $(53.22\pm0.06)$~d and a $(10.44\pm0.04)\%$ branching ratio for decay to the 478-keV daughter state~\cite{tunl}. 

\subsubsection{Online (prompt) Ge detector}
\label{online}
\begin{figure}
\includegraphics[width=0.45\textwidth]{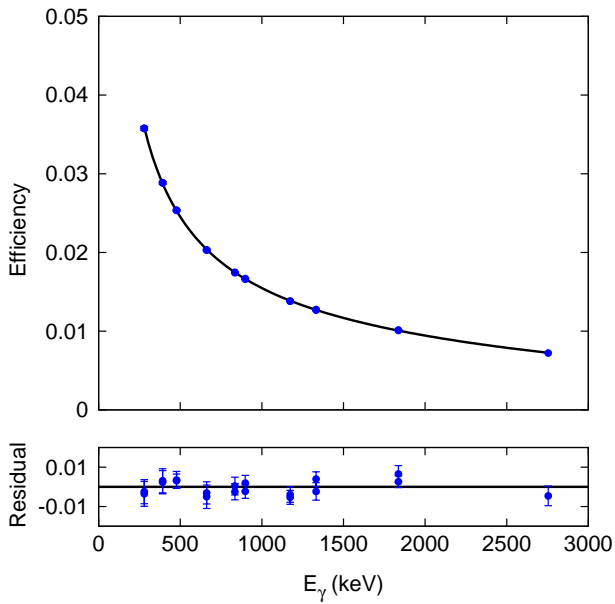}
\caption{(Color online) Absolute online Ge photopeak detection efficiency vs.~$E_{\gamma}$.  Solid points -- measurements.  Error bars are statistical.  Curve -- fit to the data of the function $\epsilon(E_{\gamma}) = aE_{\gamma}^b(1+cE_{\gamma}^2)$, with $a,b$ and $c$ parameters.  Bottom panel: fractional fit residuals.}
\label{ge-eff}
\end{figure}

Online Ge detector efficiency measurements in the range $E_{\gamma}$ = 280 to 2754 keV were made using the $^{24}$Na, $^{54}$Mn, $^{60}$Co,  $^{137}$Cs,  $^{88}$Y, $^{113}$Sn and $^{203}$Hg sources mounted in the center of the gas cell.  For the two-line sources, summing corrections of up to $(5\pm1)\%$ were made based on measured peak/total ratios and computed $\gamma - \gamma$ angular correlation coefficients.   The results, shown in Fig.~\ref{ge-eff}, determine the online detection efficiency to $\pm1.6\%$.  The position of the online detector was not exactly the same for the measurements at different bombarding energies.  We assigned an additional  $\pm1\%$ variable systematic error to its efficiency based on numerous repeat efficiency measurements.  

\begin{figure}
\includegraphics[width=0.45\textwidth]{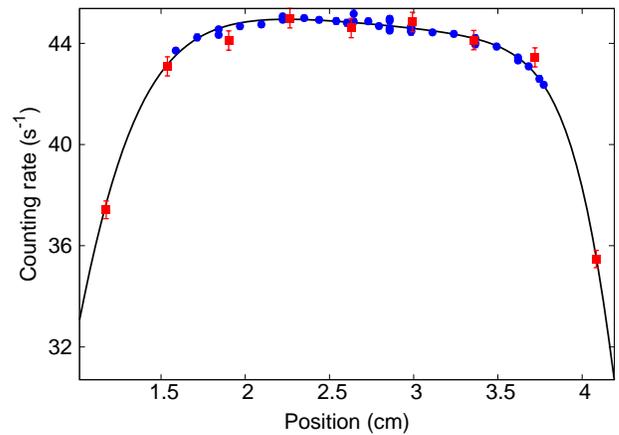}
\caption{(Color online) Solid (blue) circles: measured online Ge detector efficiency vs.~position for a  $^{137}$Cs source on the axis of the gas cell.  Solid (red) squares: {\sc Penelope} simulations.  The curve is to guide the eye.   The center of the cell is at 2.63~cm, the simulated end points are located at positions near the ends of the cell, and the position scale has an arbitrary offset.}
\label{long-eff}
\end{figure}

We made additional efficiency measurements with the sources mounted inside the cell and displaced along its axis, as shown in Fig.~\ref{long-eff}.  Increased absorption causes the efficiency to drop near the foil holder and near the back plate.  The measured dependence of efficiency on displacement is reproduced well by {\sc Penelope}~\cite{penelope} Monte Carlo efficiency simulations described below.

The $\gamma_0$ yield was obtained by summing the counts in a background-subtracted window around the full-energy peak, and the 429-keV yield was determined by fitting the peak shape with the same function used for fitting the source lines and the 478-keV line.

\subsubsection{{\sc Penelope} Monte Carlo efficiency calculations}
\label{pen-eff}
Prompt $\gamma$-rays were emitted from an extended cylindrical volume inside the gas cell.  The prompt $\gamma_0$ energy distribution was broadened by beam energy loss in the gas and by the Doppler shift, and as a result the broad $\gamma_0$ peak contained a $2 - 3$\% Compton scattering component.  In order to account accurately for these effects we carried out detailed efficiency simulations using  the {\sc Penelope} computer code~\cite{penelope}.

\begin{figure}
\includegraphics[width=0.35\textwidth]{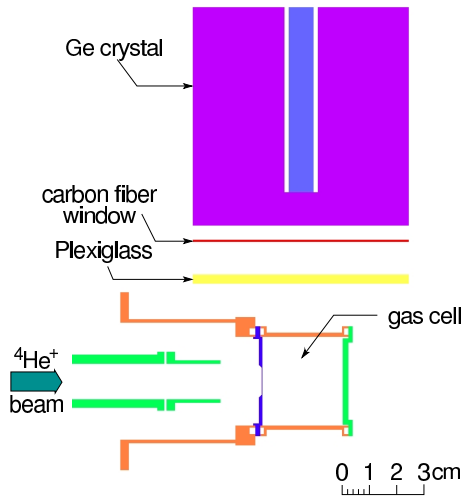}
\caption{(Color online) Gas cell and detector geometry used in {\sc Penelope} simulations.  The beam enters from the left, passing through the collimator mounting tube and electron suppressor.  The Ge detector is shown with its carbon fiber window and Plexiglass absorber.}
\label{pen-geo}
\end{figure}

\begin{figure}
\includegraphics[width=0.45\textwidth]{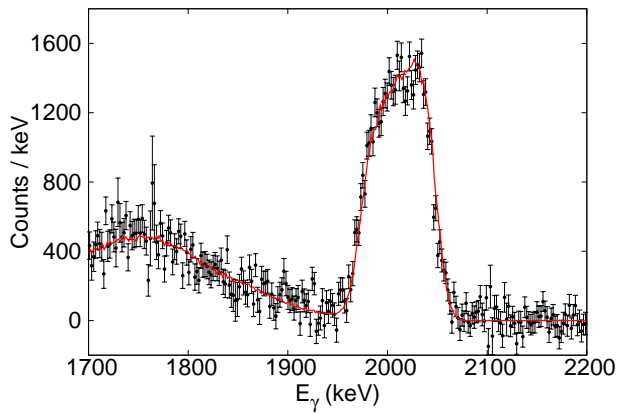}
\caption{(Color online) $E_{\rm c.m.} = 0.43$~MeV ($E_{\alpha} = 1.7$~MeV) prompt background-subtracted $\gamma$-spectrum (points) and {\sc Penelope} simulation (curve) in the region of the $\gamma_0$ peak.}
\label{pen-spectrum}
\end{figure}

Fig.~\ref{pen-geo} shows the detector and gas cell geometry used in the {\sc Penelope} calculations.  Since we needed simulated $\gamma$-ray spectra only for energies near the photopeak, we included only materials that resulted in line-of-sight absorption and/or small angle forward scattering between the cell and the detector.  Detector properties were taken from manufacturer's specifications~\cite{canberra}.   Efficiency simulations for the calibration source energies were typically $(5-7)$\% higher than our measurements for $E_{\gamma}=0.4-3.0$~MeV.

We modeled the $\gamma_0$ detection efficiency using {\sc trim} to calculate the energy and position dependence of $\gamma$ rays emitted from the gas cell for each bombarding energy.  The measured Ni foil thickness, $^3$He gas pressure and accelerator beam energy were inputs for these calculations.  The variation of cross section with energy was included by assuming a constant $S$-factor over the range of energy loss in the cell.   The result of these calculations served as input for the {\sc Penelope} simulation.  In general the simulated lineshapes agreed very well with the data, as can be seen for the $E_{\rm c.m.} = 0.43$~MeV case shown in Fig.~\ref{pen-spectrum}.

In order to minimize our sensitivity to calculations, we used {\sc Penelope} efficiency simulations only for ratios that are close to unity.  We determined the  $\gamma_0$ and $\gamma_{429}$ detection efficiencies $\epsilon(E_{\gamma})$ from the relation
\begin{equation}
\epsilon(E_{\gamma}) = \epsilon_p(E_{\gamma}) \frac{\epsilon_d'(E_{\gamma})}{\epsilon_p'(E_{\gamma})}, 
\label{effratio}
\end{equation}
where $\epsilon_p(E_{\gamma})$ is the point-source efficiency measured at the center of the cell, $\epsilon_d'(E_{\gamma})$ is the {\sc Penelope} distributed-source efficiency, $\epsilon_p'(E_{\gamma})$ is the {\sc Penelope} point-source efficiency and $E_{\gamma}$ is the mean $\gamma$-ray energy.
 Small (several keV) inaccuracies in the simulated $\gamma_0$ energy distribution were taken into account by adjusting the window used to sum the simulated yield by the difference in the measured and simulated  $\gamma_0$ centroids.    The calculated {\sc Penelope} efficiency ratios were all within  4\% of unity.   This, together with the fact that  the measured and simulated point-source efficiencies agree well as described above, suggests a  systematic uncertainty on the simulated efficiency ratio which is less than the typical simulation statistical uncertainty of $\approx0.4\%$.

\subsection{Beam current integration}
We determined the number of $\alpha$ particles entering the gas cell by integrating the electrical current measured on the gas cell, including foil holder and cell support tube, which acted as a Faraday cup as shown in Fig.~\ref{cell}.   Beam collimation and alignment ensured that the beam passed cleanly through the foil holder aperture. 
Secondary electron loss from the foil was suppressed with $-600$~V applied to the electron suppressor.  The (negative) current on the last (8~mm) collimator just upstream of the suppressor was always less than 1\% in magnitude compared to the (positive) cell current.  The current integrator was checked to a precision of better than 1\% with two different precision DC current sources and with a square wave generator.

We also made measurements without foil or gas, in which we compared the integrated beam current to the integrated beam power determined with a calorimeter suitable for relative measurements.  We used both proton and $^4$He$^+$ beams at 1.00~MeV, and we compared a geometry close to that used for the cross section measurements with a different Faraday cup geometry.  In all these comparisons, the ratio of the number of beam particles determined from the integrated beam current and from the integrated beam power was the same within the measurement accuracy of $\pm  1\%$.


A potential concern is the possiblity of He$^+$ charge-changing collisions in the residual gas of the beamline upstream of the target.  Pickup is negligible compared to stripping~\cite{sataka}, and the strong defocussing effect of the beam transport, which would transmit less than 1\% of a He$^{++}$ beam when tuned for He$^{+}$ ions, indicates that only those He$^{++}$ ions produced in the 2-m  beamline section downstream of the last quadrupole lens may reach the target cell.   In this region the beamline pressure was $ < 2 \times 10^{-6}$ torr.  The stripping cross section on O$_2$ and N$_2$ shows a broad maximum at $E_{\alpha} \approx$ 1.0 MeV and a peak cross section of $\approx 1.6\times 10^{-16}$ cm$^2$~\cite{sataka}.  These parameters imply the fraction of  He$^{++}$ ions reaching the target  was $< $0.2\%.  The rastering  would reduce the He$^{++}$ fraction by an additional factor of $\approx$4; hence He$^+$ charge-changing collisions were not an important concern for our measurements.

\subsection{Target cell gas and foil properties}
The nickel entrance foils were 1-$\mu$m nominal thickness, glued onto the nickel foil holders.  Fresh foils were used for most bombarding energies.  Foil thicknesses were measured to $\pm 1\%$  using  the 3.2-MeV $\alpha$ particles from $^{148}$Gd.  Most foils leaked He at a slow rate, and the cell pressure was ``topped off" in order to maintain a constant average pressure over time.  Pressure variations over the course of a 1-hour irradiation were $<6\%$.    The increase in cell depth due to foil bulge was measured to be $0.3-0.4$~mm averaged over the $7-8$-mm beam diameter.

We used $^3$He gas with 99.999\% (99.99\%) chemical (isotopic) purity, and naturally occuring He gas with 99.9999\% chemical purity.   The cell was flushed with $^3$He (or $^4$He) before and during the cross section measurements, which were carried out at nominal pressures of 200 torr (two highest energies) and 100 torr.  Gas purity was measured \textit{in situ} after some time in the cell, using a residual gas analyzer and found to $>99\%$ pure.  Gas pressure was measured with a Baratron~\cite{baratron}, whose precision was checked with a mercury manometer to better than 1\% at $100-200$~torr and was recorded in 1-s time intervals with an ADC (Analog-to-Digital Converter).   Cell temperature was measured with an insulated probe attached to the cell wall and recorded at the beginning and end of each one-hour run.

\subsection{Contaminant \boldmath{$^7$}Be production}
\label{contaminant}
As Hilgemeier \emph{et al.}~\cite{hilg} pointed out, $^7$Be may be produced by a background process, \emph{e.g.}~$^6$Li(\emph{d,n})$^7$Be or $^{10}$B(\emph{p},$\alpha)^7$Be if there is both a (proton or deuteron) contamination in the $^4$He beam and a ($^6$Li or $^{10}$B) contaminant in the gas cell, \emph{e.g.}~the foil or beam stop.  This concern was apparently overlooked in earlier $^3$He + $^4$He $\rightarrow$ $^7$Be activity measurements.  We guarded against this possibility by counting the stoppers from the \fourplusfour\ irradiations for 5 of the 8 bombarding energies.  We obtained $1 - 2\%$ upper limits ($1\sigma$) on contaminant $^7$Be production from each of these measurements.

In order to understand the possible magnitude of this problem in older  $^3$He + $^4$He $\rightarrow$ $^7$Be activity measurements, we made a study of beam and target stopper impurities.  Using our 90$^{\circ}$ analyzing magnet and a 3-MeV $^4$He$^+$ beam, we found a D$_2^+$ and/or DH$_2^+$ satellite beam with intensity $0.1 - 1\%$ (depending on source preparation) at slightly higher rigidity than the $^4$He$^+$ beam.  Accelerator voltage instabilities, \emph{e.g.\@} sparks, could allow this beam to pass through the analyzing magnet slits with a non-zero duty factor.  We then bombarded various Co, Ni, Cu, Nb, Ta, Pt and Au materials selected for good chemical purity ($> 99.9$\%) with 0.75-MeV protons and 1.5-MeV deuterons (the energies that would result in the above example) and measured the $^7$Be activity produced~\cite{annrep}.   Our results suggest that  contaminant $^7$Be production in the older activity experiments was probably not significant.  
We note that the Osborne \emph{et al.}~activity measurements~\cite{osbo} were made with a Tandem accelerator in which the beam was stripped in the terminal and hence was unlikely to contain molecular impurities.  We also note that a $^3$He beam experiment is more dangerous since $^7$Li($^3$He,\emph{t}) and $^6$Li($^3$He,\emph{d}) reactions can lead to contaminant $^7$Be production without the need for a beam contaminant.

\subsection{Beam heating corrections}
The region of the gas cell illuminated by the beam has a higher temperature and hence lower density than the rest of the cell.  We measured this effect using narrow resonances in the $^{24}$Mg($\alpha, \gamma)^{28}$Si and $^{10}$B($\alpha, p-\gamma)^{13}$C reactions at $E_{\rm res}$= 3.20 MeV and 1.51 MeV, respectively.  We coated the gas-cell stopper with a thin layer of natural Mg or B and measured the thick-target resonance yield with and without the foil and gas.  Then, with the foil and gas present, and the beam energy adjusted halfway up the leading edge of the resonance, we measured the resonance yield as a function of the beam current.  The gas density correction factors determined from these data are shown in Table~\ref{beamheat}.  The beam heating corrections for our 200-torr $^3$He + $^4$He $\rightarrow$ $^7$Be measurements were determined by interpolating between the entries shown in Table~\ref{beamheat}, while the 100-torr $^3$He + $^4$He $\rightarrow$ $^7$Be corrections were determined by extrapolating the 100-torr measurement, scaling by the calculated gas $dE/dx -$ see Table~\ref{properties}.

\begin{table}
\caption{Beam heating density correction factors, percent per 600~nA.}
\label{beamheat}
\begin{ruledtabular}
\begin{tabular}{cccc}
Case & $E_{\rm res}$(MeV) & 100~torr & 200~torr  \\
\hline $^{24}$Mg($\alpha, \gamma)^{28}$Si & 3.20 & -- &  3.9$\pm$0.6   \\
$^{10}$B($\alpha, p-\gamma)^{13}$C & 1.51 & 5.3$\pm$0.8 & 7.8$\pm$1.0  \\
 
\end{tabular}
\end{ruledtabular}

\end{table}

\subsection{\boldmath{$\gamma_0$} anisotropy}
\label{anis}
The 429-keV $\gamma$ ray following $\gamma_1$ emission is isotropic, since it proceeds from a $J = 1/2$ state; however, $\gamma_0$ need not be.  Tombrello and Parker~\cite{tompark} calculated the angular anisotropy, which they found to be small but nonzero in our energy range.  Column 6 of Table~\ref{oursfactors} shows the  90$^{\circ}$ $\gamma_0$ anisotropy correction factors $1/(1 - a_2Q_2/2)$ calculated from~\cite{tompark}, where $a_2$ is the coefficient of the Legendre Polynomial $P_2$ and $Q_2 = 0.9$ is the angular attenuation coefficient for our geometry.  These corrections range from 0.99 to 1.04 for our energies.  Kim \emph{et al.}~\cite{kim} calculated the $\gamma_0$ angular distribution at several energies; at $E_{\rm c.m.} = 0.1$~MeV their result agrees well with~\cite{tompark} while at $E_{\rm c.m.} = 1$~MeV their result corresponds to an anisotropy correction factor 2\% larger than~\cite{tompark}. 

The angular distribution is sensitive to the phase difference between s- and d-wave capture amplitudes, which has not been measured in our energy range.  Both calculations assumed a hard-sphere extrapolation of the nuclear part from data at higher energies, where measurements are consistent at the level of 5$^{\circ}$~\cite{phase}.  We estimate that a 5$^{\circ}$ change in this phase difference would  give rise to a 2\% change in the anisotropy correction at our energies, on average.  Thus it seems reasonable to assume a $\pm$ 2\% uncertainty in this correction.  We note that Krawinkel \emph{et al.}~\cite{kraw} are the only ones to have measured the anisotropy, and their results do not constrain it at this level of precision.  
\begin{figure}
\includegraphics[width=0.45\textwidth]{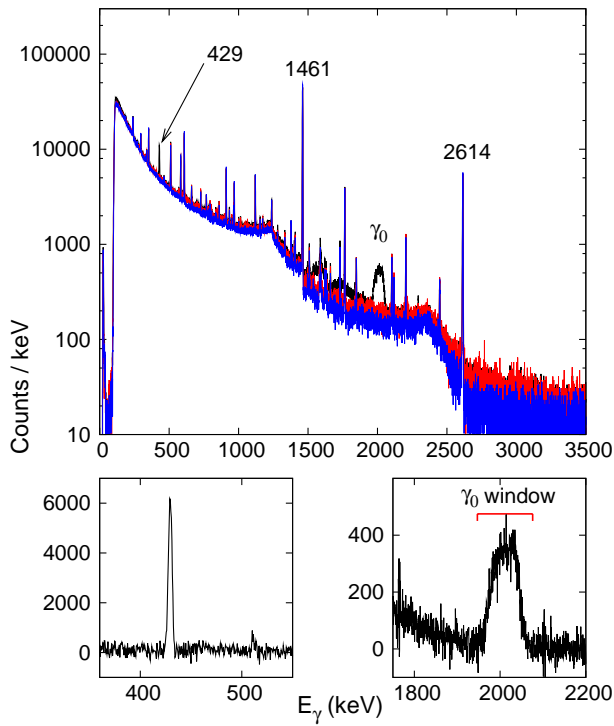}
\caption{(Color online) Prompt $\gamma$-spectra measured at  $E_{\alpha} = 1.7$~MeV ($E_{\rm c.m.} = 0.43$~MeV).  Top panel, from top to bottom: \fourplusthree\ (black), \fourplusfour\ (red), and beam-off background (blue) spectra. Bottom panels: background subtracted spectra in the region of the 429 keV line and $\gamma_0$.  The window used to determine the $\gamma_0$ yield is indicated.}
\label{1700}
\end{figure}

\begin{figure}
\includegraphics[width=0.45\textwidth]{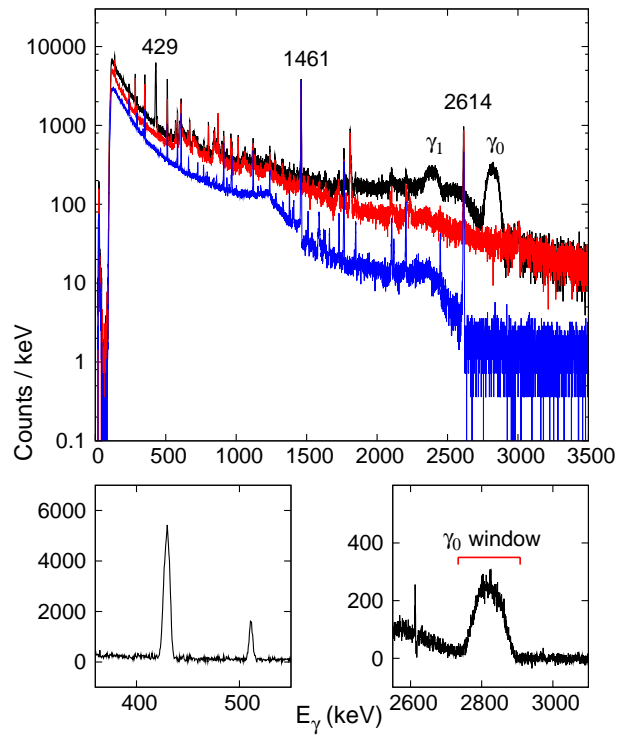}
\caption{(Color online) Prompt $\gamma$-spectra measured at $E_{\alpha} = 3.5$~MeV ($E_{\rm c.m.} = 1.23$~MeV).  The colors correspond to those used in Fig.~\ref{1700}.   See Fig.~\ref{1700} caption for additional information.}
\label{3500}
\end{figure}

\section{Results}
We measured the $^3$He + $^4$He $\rightarrow$ $^7$Be fusion cross section at eight bombarding energies from $E_{\rm c.m.} = 0.33$ to 1.23 MeV.  Some properties of these measurements are shown in Table~\ref{properties}.  At each bombarding energy we measured the \fourplusthree\ yield in a series of 1-hour runs for the total time indicated.  The rastered beam current on the cell was $420 - 500$ nA.

\fourplusfour\ background measurements were also made at each bombarding energy, for an irradiation time  $60-130\%$ of the \fourplusthree\ irradiation time.  Some background increase with time was observed for most of the runs, which we attributed mainly to the $^{13}$C($\alpha$,n)$^{16}$O reaction on carbon buildup.   \fourplusfour\ measurements were made before and after the \fourplusthree\ measurements, and summed, in order to correct, to first order, for the prompt $\gamma$-ray background increase with irradiation time.   Separate stoppers were used for the \fourplusthree\ and \fourplusfour\ irradiations so we could count the \fourplusfour\ stoppers to check for contaminant $^7$Be production (see Sec.~\ref{contaminant}).

\begin{table*}
  \caption{Conditions of the cross section measurements.}
  \label{properties}
  \begin{ruledtabular}
    \begin{tabular}{ddddcdcddd}
      \multicolumn{1}{c}{$E_{\rm c.m.}$(MeV)} 
      &    \multicolumn{1}{c}{$E_{\alpha}$(MeV)}
      &    \multicolumn{1}{c}{Foil($\mu$m)}
      &    \multicolumn{1}{c}{$\Delta E_{\rm foil}$(MeV)}
      &    \multicolumn{1}{c}{P$_{\rm gas}$(torr)\footnotemark[1]}    
      &    \multicolumn{1}{c}{$\Delta E_{\rm gas}$(MeV)}    
      &    \multicolumn{1}{c}{I$_{\rm av}$(nA)}    
      &    \multicolumn{1}{c}{BH($\%$)\footnotemark[2]}        
      &    \multicolumn{1}{c}{T$_{\rm irrad}$(hr)\footnotemark[3]}        
      &    \multicolumn{1}{c}{T$_{\rm act}$(d)\footnotemark[4]}        \\
      0.3274    &1.5    &    0.968    &    0.677    &    100    &    0.166    
      &    480    &    5.3    &    102.3    &    26.6    \\
      0.4260     &1.7    &    0.937    &    0.641    &    100    &    0.158    
      &    460    &    4.9    &    116.5    &    12.0    \\
      0.5180     &1.9    &    0.955    &    0.638    &    100    &    0.148    
      &    450    &    4.5    &    30.9    &    33.6    \\
      0.5815    &2.1    &    1.069    &    0.691    &    100    &    0.141    
      &    420    &    3.9    &    24.4    &    20.5    \\
      0.7024    &2.35    &    1.069    &    0.665    &    100    &    0.129    
      &    450    &    3.9    &    23.3    &    10.4    \\
      0.7968    &2.6    &    1.072    &    0.642    &    200    &    0.240    
      &    500    &    5.9    &    20.3    &    7.2    \\
      1.2337    &3.5    &    1.065    &    0.545    &    200    &    0.182    
      &    440    &    3.5    &    14.7    &    6.8    \\
      1.2347    &3.5    &    1.051    &    0.538    &    200    &    0.182    
      &    120/500    &    \multicolumn{1}{c}{1.2/4.9}    &    41.8  &  31.8\\
    \end{tabular}
  \end{ruledtabular}
  \footnotetext[1]{Nominal gas pressure.}
  \footnotetext[2]{Average beam heating correction. }
  \footnotetext[3]{\fourplusthree\ irradiation time.}
  \footnotetext[4]{Activity counting time.}
\end{table*}

The $\gamma$-ray spectra measured at $E_{\rm c.m.} = 1.23$ and 0.43 MeV are shown in Figs.~\ref{1700} and \ref{3500}.  For most of the measurements the beam-related background (per $\mu$C) determined from \fourplusfour\ was statistically consistent with the background observed above the $\gamma_0$ peak in the \fourplusthree\ spectra.  Even so,  we adopted a more conservative procedure of determining the \fourplusthree\ background by normalizing the (beam-related part of the) \fourplusfour\ spectrum to a region above the $\gamma_0$ peak in the \fourplusthree\ spectrum.  The \fourplusfour\ spectra shown in the top panels of these figures are shown with this normalization.  The lower panels in Figs.~\ref{1700} and \ref{3500} show the background-subtracted prompt spectra in the region of the $\gamma_0$ and 429-keV peaks, and the window used to sum the $\gamma_0$ counts.   Sharp features in the subtracted spectra near 2614 and 1765 keV originate from large  background peaks that have been subtracted imperfectly (due to small gain shifts) and have small net areas.

In our $E_{\rm c.m.}= 0.80$ MeV data we observed a large carbon buildup.  This allowed us to check that the beam-related backgrounds due to carbon buildup and the background present before buildup had the same shape in the region of interest.  Thus our method of background determination under the prompt $\gamma_0$ peak appears to be robust.  

\begin{figure}
\includegraphics[width=0.45\textwidth]{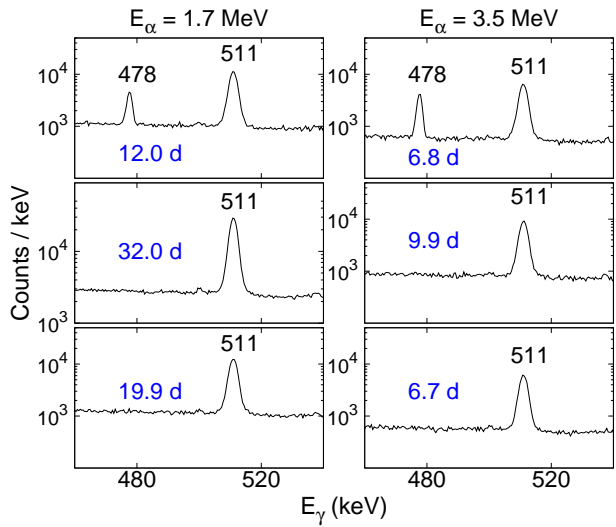}
\caption{(Color online) $E_{\rm c.m.} = 1.23$ and 0.43 MeV ($E_{\alpha}$ = 3.5 and 1.7 MeV) offline activity spectra.   Top: \fourplusthree\ irradiation.  Middle: \fourplusfour\ irradiation.  Bottom: source-out background.  Counting periods (in d) are shown.}
\label{activityspectra}
\end{figure}

The $E_{\rm c.m.}$= 1.23 and 0.43 MeV activity spectra are shown in Fig.~\ref{activityspectra}.  Neither the source-out background (measured with a Cu stopper that had not been irradiated) nor the \fourplusfour\ stopper measurements showed any sign of a 478-keV peak.

We determined the mean reaction energy from the $\gamma_0$ centroid.   The advantage of this method is that it does not depend on knowledge of the beam energy or the energy loss in the foil or gas, and it is determined directly from the events used to determine the cross section.  However, the broad $\gamma_0$ peak contains a small  forward-angle Compton scattering component in addition to the majority of events in which the full $\gamma$-ray energy is deposited in the Ge.  The $\gamma_0$ centroid must be corrected for this component in order to obtain the mean reaction energy.  (Note that the Compton contribution to the $\gamma_0$ detection efficiency is included in the {\sc Penelope} ratio factor that appears in  Eq.~\ref{effratio}).  We computed centroid corrections of 1.2 $\pm$ 0.1 and 1.9 $\pm$ 0.1 keV for the  $E_{\rm c.m.}$=  0.33 and 1.23 MeV cases using the measured detector response for $E_{\gamma}$= 1836 and 2754 keV, and we interpolated linearly to obtain the corrections for the other reaction energies.   These corrections shift our inferred $S$-factors down by 1.2\% (0.9\%)  for the $E_{\rm c.m.}$=  0.33 (0.43) MeV cases and less at the higher energies.

We also corrected for a small net positive Doppler shift in the $\gamma_0$ centroid.  This resulted primarily because, on average, the $\gamma$-rays were emitted somewhat ($\leq 2$~mm) upstream of the center of the cell.   The calculated correction is 0.35 (0.2) keV for the $E_{\rm c.m.} = 0.33 (0.43)$~MeV cases, and has the opposite sign to the Compton correction discussed above.  All of our quoted $E_{\rm c.m.}$ values and $S$-factors include these corrections.

\begin{table*}
\caption{Our $S$-factors, prompt $\gamma$-ray branching ratios, and calculated $\gamma_0$ anisotropy.}
\label{oursfactors}
\begin{ruledtabular}
\begin{tabular}{cccccc}
$E_{\rm c.m.}$(MeV)		&		$S_{\rm act}$(keV\,b)\footnotemark[1] 		&		$S_{\rm prompt}$	(keV\,b)\footnotemark[1]$^,$\footnotemark[2]	&		$S_{\rm prompt}/S_{\rm act}$\footnotemark[2]		&		$\gamma_{429}/\gamma_0$\footnotemark[2]		&		Anisotropy\footnotemark[3]	\\
\hline
0.3274	$\pm$	0.0013	&	0.495	$\pm$	0.015	&	0.492	$\pm$	0.014	&	0.994	$\pm$	0.042	&	0.410	$\pm$	0.023	&	0.988	\\
0.4260	$\pm$	0.0004	&	0.458	$\pm$	0.010	&	0.438	$\pm$	0.006	&	0.956	$\pm$	0.024	&	0.405	$\pm$	0.009	&	1.002	\\
0.5180	$\pm$	0.0005	&	0.440	$\pm$	0.010	&	0.421	$\pm$	0.007	&	0.957	$\pm$	0.026	&	0.394	$\pm$	0.009	&	1.013	\\
0.5815	$\pm$	0.0008	&	0.400	$\pm$	0.011	&	0.398	$\pm$	0.008	&	0.995	$\pm$	0.035	&	0.422	$\pm$	0.013	&	1.019	\\
0.7024	$\pm$	0.0006	&	0.375	$\pm$	0.010	&	0.382	$\pm$	0.006	&	1.020	$\pm$	0.031	&	0.424	$\pm$	0.010	&	1.030	\\
0.7968	$\pm$	0.0003	&	0.363	$\pm$	0.007	&	0.371	$\pm$	0.004	&	1.022	$\pm$	0.023	&	0.427	$\pm$	0.005	&	1.036	\\
1.2337	$\pm$	0.0003	&	0.330	$\pm$	0.006	&	0.327	$\pm$	0.004	&	0.990	$\pm$	0.021	&	0.439	$\pm$	0.006	&	1.040	\\
1.2347	$\pm$	0.0003	&	0.324	$\pm$	0.006	&	0.340	$\pm$	0.004	&	1.050	$\pm$	0.023	&	0.443	$\pm$	0.007	&	1.040	\\
\end{tabular}
\end{ruledtabular}
\footnotetext[1]{Uncertainties include statistics, $\pm 1\%$ variable systematic and the contribution from  the $E_{\rm c.m.}$ uncertainty.}
\footnotetext[2]{Assuming $\gamma_0$ isotropic. }
\footnotetext[3]{$\gamma_0$ anisotropy correction $1/(1 - a_2Q_2/2)$ calculated from~\protect\cite{tompark}.}
\end{table*}

As an additional check on beam current integration, we remeasured the cross section at $E_{\rm c.m.} = 1.23$~MeV using 4-mm instead of 7-mm diameter beam-defining apertures.  This further minimized the secondary electron currents on the cleanup collimators.  We determined the beam heating correction for this collimation geometry directly from $^3$He + $^4$He $\rightarrow$ $^7$Be data measured at 120 and 500~nA.  The good agreement between the activity $S$-factors deduced from these two $E_{\rm c.m.} = 1.23$~MeV cross section measurements provides additional evidence that our beam current integration and beam heating estimates are reliable.  The prompt $S$-factors for the two 1.23 MeV measurements agree less well; however, here we measured the \fourplusfour\ beam-related background only at 500 nA.

\subsection{\boldmath{$S$}-factors}
Our measured $S$-factors are shown in Table~\ref{oursfactors}.  The mean center-of-mass reaction energy was determined by subtracting the 1586.1 $\pm$ 0.1 keV Q-value~\cite{wapstra} from the $\gamma_0$ centroid.  Cross sections were converted to $S$-factors using the relation
\begin{equation}
{S}(E_{\rm c.m.}) =  \sigma( E_{\rm c.m.})E_{\rm c.m.}e^{(E_G/E_{\rm c.m.})^{1/2}},
\label{fE0}
\end{equation}
with $E_G^{1/2}$ = 164.13 keV$^{1/2}$.  The tabulated uncertainties in the $S$-factors and their ratios include statistics, the $\pm 1\%$ variable systematic uncertainties shown in Table~\ref{syserrors}, and the contribution of the uncertainty in the mean reaction energy $E_{\rm c.m.}$ -- the latter is important only at the lowest bombarding energy.  We also show the experimental branching ratio $\gamma_{429}/\gamma_0$ ( = $\gamma_1/\gamma_0$) and the anisotropy calculated from~\cite{tompark} (see Sec.~\ref{anis}).  

\begin{table}
\caption{Systematic uncertainties $\Delta S(E_{\rm c.m.})/S(E_{\rm c.m.}$). 
  The variable uncertainties are assumed random between the various 
  $S(E_{\rm c.m.}$) values, while the scale factor uncertainties are the same 
  for all the values.}
\label{syserrors}
\begin{ruledtabular}
  \begin{tabular}{lccl}
    & Type\footnotemark[1] & $\Delta$S/S(\%) & \multicolumn{1}{c}{Origin} \\
    \hline
    variable: & a, p & 1.0 & source positioning \\
    \hline
    scale factor: & a, p  & 1.6 & Ge efficiency\footnotemark[2] \\
    &	p  &	1.0 & beam centering  \\
    &	p  &	1.5 & $\gamma_0$ anisotropy \\
    &	a, p  &  1.6 & current integration \\
    &	a, p  & 1.3 & gas pressure \\
    &	a, p  & 0.6 &temperature \\
    &	a, p & 0.7 & gas purity \\
    &	a, p  & 0.4 & gas cell length\footnotemark[3]	 \\
    &	a, p  & 1.0 & beam heating \\
    \hline
    total scale factor: &	a   &	3.0 \\
    &	p   &	3.5 \\
    \multicolumn{2}{r}{common} & 2.7 \\
  \end{tabular}
\end{ruledtabular}
\footnotetext[1]{Activity and/or prompt.}
\footnotetext[2]{Prompt value includes summing and simulation uncertainties.}
\footnotetext[3]{Includes foil bulge.}
\end{table}

Table~\ref{syserrors} shows the sources of systematic uncertainty and their contributions to the $S$-factor uncertainty.  The $\pm$ 1\% variable systematic uncertainties given in Table~\ref{syserrors} are included in the $S$-factor values and ratios shown in Table~\ref{oursfactors} and in the fitting of $S(E_{\rm c.m.}$) vs. $E_{\rm c.m.}$, while the systematic scale-factor uncertainty is folded with the fit uncertainty to give the total uncertainty.  The fit uncertainties include the factor $\sqrt{\chi^2/\nu}$ whenever $\chi^2/\nu > 1$.

\begin{figure}
\includegraphics[width=0.45\textwidth]{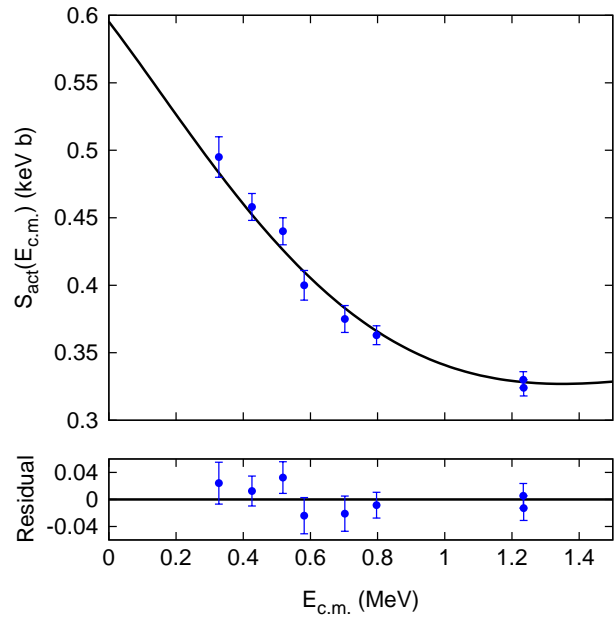}
\caption{(Color online) Solid (blue) points: $S_{\rm act}$($E_{\rm c.m.}$) vs. $E_{\rm c.m.}$.   Curve: fit of Kajino's function.  Bottom panel: fractional fit residuals.}
\label{sact}
\end{figure}

\begin{figure}
\includegraphics[width=0.45\textwidth]{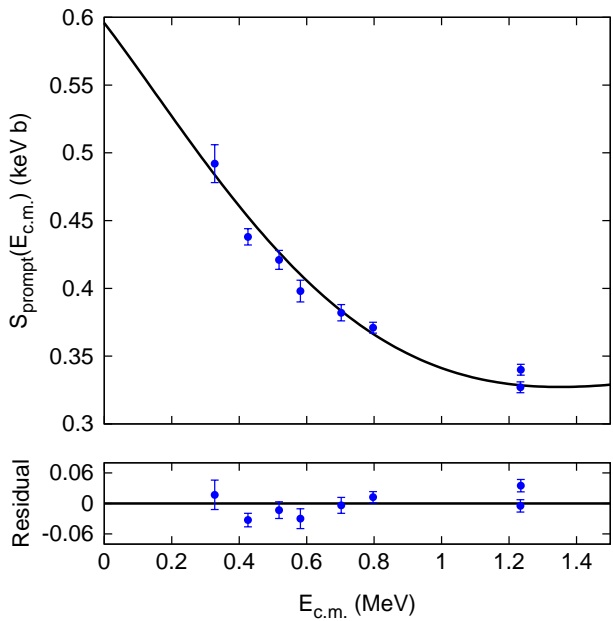}
\caption{(Color online) Solid (blue) points: $S_{\rm prompt}(E_{\rm c.m.})$ vs. $E_{\rm c.m.}$, for the isotropic $\gamma_0$ assumption.   Curve: fit of Kajino's function.  Bottom panel: fractional fit residuals.}
\label{sprompt}
\end{figure}

Fig.~\ref{sact} shows $S_{\rm act}$($E_{\rm c.m.}$) vs. $E_{\rm c.m.}$ together with a fit in which the theoretical shape given by Eq. 6 of Kajino \textit{et al.}~\cite{kajino} has been scaled to fit the data.   Fig.~\ref{sprompt} shows $S_{\rm prompt}$($E_{\rm c.m.}$)  
 based on an assumed isotropic angular distribution for $\gamma_0$ together with the Kajino fit.  In these fits the shape parameters of the fit function are held constant and only one parameter, $S(0)$, is varied.

\begin{figure}
\includegraphics[width=0.45\textwidth]{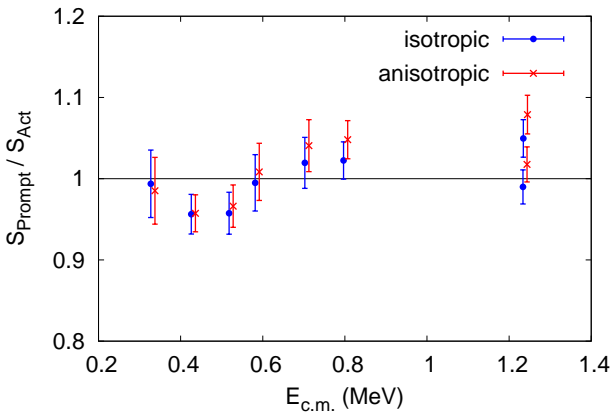}
\caption{(Color online) $S_{\rm prompt}$/S$_{\rm act}$ vs. $E_{\rm c.m.}$.   Solid (blue) circles: assuming $\gamma_0$ isotropic.  (Red) x's: $\gamma_0$ corrected for theoretical anisotropy as given in Table~\ref{oursfactors}.  The crosses have been displaced from the circles for clarity.}
\label{ratios}
\end{figure}

\begin{table}
  \caption{Our $S(0)$ values and prompt/activity ratios.}
  \label{ours0}
  \begin{ruledtabular}
    \begin{tabular}{lldd}
      \multicolumn{1}{c}{Quantity} & 
      \multicolumn{1}{c}{Value\footnotemark[1]} &  
      \multicolumn{1}{c}{$\chi^2/\nu$\footnotemark[2]} & 
      \multicolumn{1}{c}{P(\%)\footnotemark[3]} \\
      \hline
 $S_{\rm act}(0)$ & $0.595\pm0.018$~keV\,b & 0.72 & 65 \\
 $S_{\rm prompt}(0)$\footnotemark[4]  & $0.596\pm0.021$~keV\,b & 2.7 & 1 \\
 $S_{\rm prompt}/S_{\rm act}$\footnotemark[4]  & $0.999\pm0.030$ & 1.8 & 9\\
 \hline
 $S_{\rm prompt}(0)$\footnotemark[5] & $0.607\pm0.022$~keV\,b & 5.3 & 0.0004 \\ 
 $S_{\rm prompt}/S_{\rm act}$\footnotemark[5] & $1.015\pm0.032$ & 2.9 & 0.5 \\
 \hline
 Combined $S(0)$\footnotemark[6] & $0.595\pm0.018$~keV\,b &   &   \\  
    \end{tabular}
  \end{ruledtabular}
  \footnotetext[1]{Uncertainties include systematic contributions.}
  \footnotetext[2]{$\nu=7$.}
  \footnotetext[3]{P($\chi^2,\nu)$ is the confidence level of the fit.}
  \footnotetext[4]{$\gamma_0$ assumed isotropic.}
  \footnotetext[5]{$\gamma_0$ corrected by theoretical anisotropy -- see Table~\ref{oursfactors}.}
  \footnotetext[6]{Activity and isotropic prompt.}
\end{table}

Fig.~\ref{ratios} shows the ratios $S_{\rm prompt}$/$S_{\rm act}$ for both the isotropic and anisotropic assumptions for $\gamma_0$.  We fit the ratios $S_{\rm prompt}$/$S_{\rm act}$ vs. $E_{\rm c.m.}$ with a horizontal line, and we also fit our $S_{\rm prompt}$($E_{\rm c.m.}$)   values based on the anisotropic assumption.
The results of all these fits are shown in Table~\ref{ours0}, including the confidence level P($\chi^2, \nu)$ for each fit.  The fit to $S_{\rm act}$($E_{\rm c.m.}$) has a good quality.  The fit to $S_{\rm prompt}$($E_{\rm c.m.}$) is very poor quality for the anisotropic case, and although the confidence level is still low, it's much better for the isotropic assumption.  The ratio fits are also poorer for the anisotropic assumption, and acceptable for the isotropic assumption.  Thus we conclude that the $\gamma_0$ anisotropies calculated from~\cite{tompark} are not consistent with our data, while the isotropic approximation is better.  Thus our best $S(0)$ values based on the Kajino fit function are
\begin{equation}
S_{\rm act}(0) = 0.595  \pm 0.018 \hspace{0.2cm}\mbox{keV\,b},
\label{s0activity}
\end{equation}
and
\begin{equation}
S_{\rm prompt}(0) = 0.596  \pm 0.021  \hspace{0.2cm}\mbox{keV\,b.} 
\label{s0prompt}
\end{equation}
Our best value for $S(0)$ is obtained by combining the activity and prompt results given above, taking into account the 2.7\% common scale factor uncertainty given in Table~\ref{syserrors}:
\begin{equation}
S(0) = 0.595  \pm 0.018  \hspace{0.2cm}\mbox{keV\,b.} 
\label{s0}
\end{equation}

\section{Comparison with other experiments}
We found that fitting a given data set with different theories resulted in $S(0)$ values that varied typically by several percent (5\% in the case of ref.~\cite{robe}, which consists of one high-energy point).  Hence in comparisons of different experimental results, it is important to use the same fitting function -- the exceptions are the LUNA results~\cite{gyur,conf} which were determined from data at very low energy.    Since the resonating-group calculations of Kajino \emph{et al.} do a good job of reproducing other observables, and result in a reasonable theoretical value for $S(0)$, we chose this fit function.   We obtained data from the original publications, from the NACRE compilation~\cite{nacre} and from other sources (see Refs.~\cite{park,osbo}).

The results are shown in Table~\ref{alls0}.  Most differences between the earlier results shown here and in Table II of~\cite{rmp} are due to the differing fitting functions; the notable exception is the Osborne \emph{et al.}~\cite{osbo} activity result, which is significantly higher here.  This problem has been noted before --  see Kajino \textit{et al.}~\cite{kajino} Table 1 and footnotes.  We fit Osborne \emph{et al.}'s $S_{\rm act}(E_{\rm c.m.})$ values quoted in their published Table 1~\cite{osbo}, which are the same as shown in their published Fig.~9 and in Osborne's Ph.D. thesis~\cite{osbo}.   Volk et al.~\cite{volk} give insufficient detail for us to interpret their integral measurement in a manner consistent with our other analyses; hence we simply quote their published value.    
\begin{table}
\caption{$S(0)$ values from our work and from published data.}
\label{alls0}
\begin{ruledtabular}
\begin{tabular}{lllc}
 & $S(0)$\footnotemark[1]	(keV\,b) &\multicolumn{2}{c}{Reference}\\
\hline	
activity: &	0.577 $\pm$	0.035  &	Osborne et al. & \protect\cite{osbo} \\
 &	0.660  $\pm$	0.040  &	Robertson et al. & \protect\cite{robe}\\
 &	0.560  $\pm$	0.030\footnotemark[2]  &	Volk et al. & \protect\cite{volk}\\
 &	0.546\footnotemark[3]  $\pm$	0.020  &	Nara Singh et al. & \protect\cite{nara}\\
 &	0.545  $\pm$	0.017  &	Gyurky et al. & \protect\cite{gyur}\\

 &	0.595  $\pm$	0.018  &	Present work  &\\
\hline
prompt:	&  0.481  $\pm$	0.053  &	Parker et al. & \protect\cite{park}\\
 &	0.579   $\pm$	0.07\footnotemark[4]  &	Nagatani et al. & \protect\cite{naga}\\
 &	0.449   $\pm$	0.06\footnotemark[5]  &	Krawinkel et al. & \protect\cite{kraw} \\
 &	0.522   $\pm$	0.03  &	Osborne et al. & \protect\cite{osbo}\\
 &	0.478   $\pm$	0.04  &	Alexander et al. & \protect\cite{alex}\\
 &	0.542   $\pm$	0.03  &	Hilgemeier et al. & \protect\cite{hilg}\\
 & 	0.560 	$\pm$	0.021 & Confortola et al. & \protect\cite{conf} \\
 &	0.596   $\pm$	0.021  &	Present work \\
\hline
total:\footnotemark[6] & 0.560 $\pm$	0.017\footnotemark[2] & Confortola et al. & \protect\cite{conf} \\
 &	0.595   $\pm$	0.018  &	Present work \\
\end{tabular}
\end{ruledtabular}
\footnotetext[1]{From our fits of Kajino's function to published $S(E)$ values, with uncertainties as quoted by authors except where noted.}
\footnotetext[2]{Value and uncertainty as quoted by authors.}
\footnotetext[3] {Scaled by 1.008 to account for the different $^7$Be decay BR assumed.}
\footnotetext[4]{Uncertainty as quoted in \protect\cite{rmp}.}
\footnotetext[5]{Value corrected by x1.4, and uncertainty as given in \protect\cite{hilg}.}
\footnotetext[6]{Combined activity and prompt.}
\end{table}

\begin{figure}
\includegraphics[ angle=90, width=0.48\textwidth ]{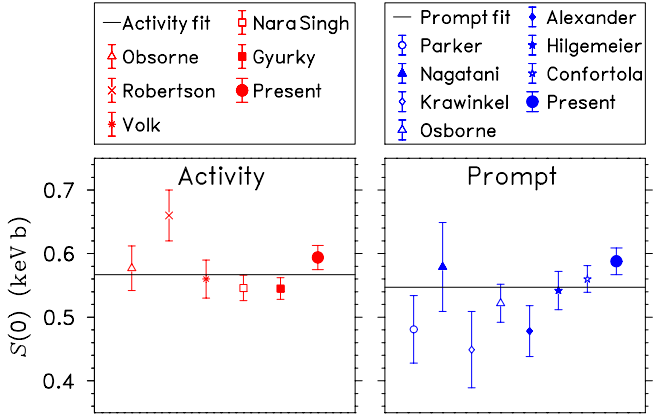} 
\caption{(Color online) Left panel, (red) points: $S_{\rm act}$(0).  Right panel,  (blue) points, $S_{\rm prompt}$(0) .  The horizontal lines are the fits to the activity values (left panel) and prompt values (right panel) as given in Table~\ref{s0comparisons}.}
\label{s0-fig}
\end{figure}

The activity and prompt  $S(0)$ values from Table~\ref{alls0} are shown in Fig.~\ref{s0-fig} together with separate fits to  the activity values  and to the prompt values.  These and other fit results  are shown in Table~\ref{s0comparisons}.  Our contaminant study reported in Sec.~\ref{contaminant} suggests earlier activity results probably did not suffer from contaminant $^7$Be production.  Earlier theoretical work showed that processes other than single-photon emission, such as E0 emission, should be negligible~\cite{snover}, while our measured prompt/activity ratios imply an experimental uppper limit on E0 emission of $\approx$2\%.  Thus we see no reason why older activity measurements, as a group, should be flawed.   However, as shown in Table~\ref{s0comparisons}, the fit to all activity values has a poor $\chi^2/\nu$ and a poor confidence level, though an acceptable quality fit is obtained if the Robertson \emph{et al.} datum \cite{robe} is removed.

\begin{figure}
\includegraphics[angle=90, width=0.48\textwidth]{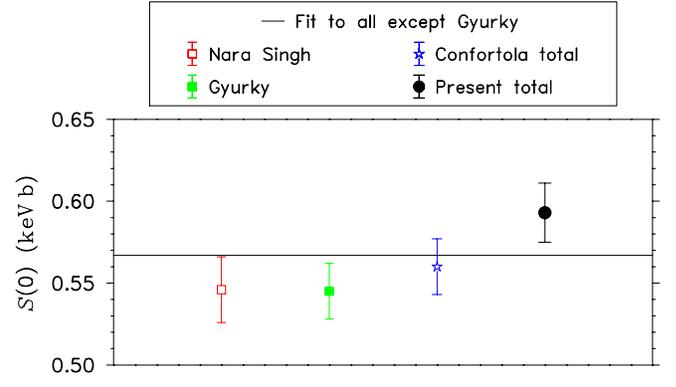}
\caption{(Color online) Recently measured $S(0)$ values.  Points, from left to right: Ref.~\protect\cite{nara} and Ref.~\protect\cite{gyur} activity values, Ref.~\protect\cite{conf} and present activity+prompt values.   Horizontal line -- fit to Ref.~\protect\cite{nara},  Ref.~\protect\cite{conf} total and present total as given in Table~\ref{s0comparisons}.}
\label{s0-recent-fig}
\end{figure}

Comparing recent measurements, our total (prompt+activity) $S(0)$ agrees with that of Confortola \emph{et al.}~\cite{conf} (see Table~\ref{alls0}), and is higher than the activity results of refs.~\cite{nara,gyur}, as shown in Fig.~\ref{s0-recent-fig}.  Combining our total value with that of \cite{conf} and with the activity result of \cite{nara} yields a minimally acceptable fit; substituting the activity result of \cite{gyur} for the \cite{conf} total value yields a similar result though the fit quality is not as good -- see Table~\ref{s0comparisons}.  Note that we do not include the results of both \cite{gyur} and \cite{conf} in the same fit since these values are presumably highly correlated, and the degree of correlation is not specified.  We also did not derive an activity value for $S(0)$ from~\cite{conf} for the same reason.

\begin{table}
  \caption{$S(0)$ comparisons.}
  \label{s0comparisons}
  \begin{ruledtabular}
    \begin{tabular}{lcdrd}
      \multicolumn{1}{c}{Data} & S(0)\footnotemark[1] & 
      \multicolumn{1}{c}{$\chi^2/\nu$} & \multicolumn{1}{c}{$\nu$} & 
      \multicolumn{1}{c}{P(\%)\footnotemark[2]} \\
      \hline 
      all activity & $0.568\pm0.014$ & 2.1 & 5  & 6  \\
      all activity except \protect\cite{robe} & $0.563\pm0.011$ & 1.3 & 4 & 27 \\
      all prompt & $0.549\pm0.016$ & 2.0 & 7  & 5   \\
      all\footnotemark[3] & $0.559\pm0.012$ & 2.1 & 11  & 2   \\
      present tot  + \protect\cite{conf} tot + \protect\cite{nara}    
      & $0.568\pm0.014$ & 1.8 & 2 & 16    \\
      present tot +  \protect\cite{gyur} +  \protect\cite{nara}   
      & $0.562\pm0.017$ & 2.5 & 2 & 8   \\
    \end{tabular}
  \end{ruledtabular}
  \footnotetext[1]{All fit uncertainties contain the factor $\sqrt{\chi^2/\nu}$}
  \footnotetext[2]{P($\chi^2,\nu)$.}
  \footnotetext[3]{All except the last 2 prompt values and the last 2 activity values in Table~\ref{alls0}.}
\end{table}

\section{Conclusion}
We have presented new results for the $^3$He + $^4$He $\rightarrow$ $^7$Be $S$-factor based on measurements of the prompt $\gamma$-rays and the $^7$Be activity produced in the same irradiation.  We find good agreement between the two methods, and a combined result 
\begin{equation}
S(0) = 0.595  \pm 0.018  \hspace{0.2cm}\mbox{keV\,b,} 
\end{equation}
based on fits of the theory of Kajino \emph{et al.}~\cite{kajino} to our data. 
 
The lack of good agreement between the various $S(0)$ values precludes a meaningful determination of a ``best" value. 
However, the recent, precise $S(0)$ determinations (ours and refs.~\cite{nara,gyur,conf}) all lie between 0.53 and 0.61 keV\,b, suggesting that the true value of $S(0)$ also lies in this range.  Comparing to older recommendations made before the recent measurements, this range is is well inside the range recommended by NACRE~\cite{nacre}, $0.54 \pm 0.09$ keV\,b, and somewhat higher than the ranges recommended by~\cite{rmp}, $0.53 \pm 0.05$ keV\,b, and by~\cite{desc}, $0.51 \pm 0.04$ keV\,b.

Further progress will probably depend on new measurements.  It may be important to note that a new measurement using a recoil separator to detect the $^7$Be nuclei, previewed in~\cite{erna}, is in good agreement with the present results in the region of overlap near $E_{\rm c.m.} = 1.2$~MeV.

\section{Impact on solar model calculations and big bang nucleosynthesis }

Our value for $S(0)$ quoted above is 12\% larger than the accepted value quoted in~\cite{rmp}.   What would be the consequence of an upward adjustment of this size for solar model neutrino flux calculations and Big Bang Nucleosynthesis (BBN)?   

It would increase the calculated Standard Solar Model (SSM) neutrino fluxes from decay of $^7$Be and $^8$B in the sun by approximately 10\%~\cite{bahcall}.  Using old (new) heavy element abundances, Bahcall \emph{et al.}~\cite{bahcall2} quote 1.09 (0.87) for the ratio of the calculated SSM to measured $^8$B neutrino flux, where the measured flux has $\pm$ 9\% uncertainty~\cite{ahmed} and the SSM flux has $\pm$ 16\%  uncertainty (both 1$\sigma$).   Thus a 10\% upward adjustment of these calculated $^8$B fluxes would not lead in either case to a significant disagreement with the measured flux, based on these uncertainties.

The uncertainty in the $^3$He + $^4$He $\rightarrow$ $^7$Be $S(0)$ is one of the largest uncertainties that contribute to the SSM neutrino flux from  $^7$Be decay in the sun~\cite{bahcall}; hence an eventual reduction in this uncertainty will reduce significantly the uncertainty in the calculated SSM flux that the BOREXINO~\cite{bor} and KamLAND~\cite{kam} experiments are designed to observe. 

The $^7$Li problem in Big Bang Nucleosynthesis  is well-known and long-standing: $^7$Li, which results mainly from decay of $^7$Be produced by the $^3$He + $^4$He $\rightarrow$ $^7$Be reaction, is overpredicted by BBN calculations~\cite{bbn}.  Thus an increase in the $^3$He + $^4$He $\rightarrow$ $^7$Be reaction rate at BBN energies would worsen this discrepancy.

\begin{acknowledgments} 

We thank the CENPA technical staff, particularly G. C. Harper and D. I. Will for their support,   R. Abel, M. K. Bacrania, A. M. Crisp, J. D. Lowrey, K. Michnicki, P. Peplowski and J. Sibille for their help, P. D. Parker for helpful comments and the U.S. DOE, Grant No. No. DE-FG02-97ER41020 for financial support.  

\end{acknowledgments}

\end{document}